\documentstyle{article}

\begin{document}

\begin{center}
{\large {\bf Modifications of Schr\"{o}dinger's Equation Complying with}}

{\large {\bf the Effect of Earth's Rotation on Quantum Energy in Atoms}}

{\large {\bf and with the Electromagnetic Force}}

$\ $

{\sf Ching-Chuan Su}

Department of Electrical Engineering

National Tsinghua University

Hsinchu, Taiwan

$\ $
\end{center}

\noindent {\bf Abstract} -- Recently, we have presented a local-ether wave
equation incorporating a nature frequency and the electric scalar potential,
from which the speed-dependences in the angular frequency and wavelength of
matter wave, in the mass of particle, and in the energy of quantum states
are derived. These relations look like the postulates of de Broglie and the
Lorentz mass-variation law, except that the particle speed is referred
specifically to a geocentric inertial frame and hence incorporates earth's
rotation for earthbound particles. Further, the wave equation is extended by
connecting the scalar potential to the augmentation operator which is
associated with a velocity difference between involved particles. Then the
electromagnetic force law is derived, which under some ordinary conditions
reduces to the modified Lorentz force law. In this investigation, the
interaction of atoms with electromagnetic radiation is explored. Then it is
shown that the time evolution equation derived from the wave equation is
substantially identical to Schr\"{o}dinger's equation incorporating the
vector potential, if the latter is observed in the atom frame and if the
source generating the vector potential is electrically neutralized, as in
common practice.

$\ $

$\ $

\noindent {\large {\bf 1. Introduction}}

Recently, we have presented a wave equation which incorporates a natural
frequency $\omega _{0}$ and the electric scalar potential and is proposed to
govern the matter wave associated with a charged particle [1]. For a
harmonic-like wavefunction, the wave equation leads to a first-order time
evolution equation similar to Schr\"{o}dinger's equation. Then it has been
found that the energies of quantum states in an atom decrease with the atom
speed by the famous Lorentz speed-dependent mass-variation factor. The
propagation of matter wave as well as electromagnetic wave is supposed to
follow the local-ether model [2]. Accordingly, the position vectors, time
derivatives, and velocities in this wave equation are all referred
specifically to an ECI (earth-centered inertial) frame for earthbound
phenomena. Thus the atom speed incorporates the linear velocity due to
earth's rotation. Consequently, the quantum state energy and the transition
frequency of an earthbound atom depend on earth's rotation, but are entirely
independent of earth's orbital motion around the Sun or others. This effect
of earth's rotation on atomic quantum properties has been used to account
for the east-west directional anisotropy in the atomic clock rate which in
turn has been demonstrated in the Hafele-Keating experiment with
circumnavigation clocks. It also accounts for the synchronism and the
clock-rate adjustment in GPS (global positioning system) and the spatial
isotropy in the Hughes-Drever experiment [1].

Furthermore, from the local-ether evolution equation, the velocity and then
the acceleration of a charged particle under the influence of the electric
scalar potential have been derived. Thus the electrostatic force is derived
in conjunction with the consequence that the natural frequency $\omega _{0}$
is related to the rest mass $m_{0}$ of the particle in the familiar form of $%
m_{0}=\hbar \omega _{0}/c^{2}$ [1]. Further, the local-ether wave equation
is extended in such a way that the electric scalar potential is made to
connect to the augmentation operator which in turn is associated with the
momentum operator and the velocity of source practices. Then the
electromagnetic force law is derived. Under the low-speed condition, this
law reduces to the modified Lorentz force law [3].

In this investigation, we derive the effect of the augmentation operator in
the interaction of atoms with electromagnetic radiation. Then the relation
between the local-ether evolution equation and Schr\"{o}dinger's equation
incorporating the vector potential is explored.

$\ $

\noindent {\large {\bf 2. Schr\"{o}dinger's Equation with Vector Potential}}

To begin with, we review the famous Schr\"{o}dinger's equation and its
consequences. It is well known that in the presence of the scalar potential $%
\Phi $ and the vector potential ${\bf A}$, Schr\"{o}dinger's equation reads 
$$
i\hbar \frac{\partial }{\partial t}\psi ({\bf r},t)=\frac{1}{2m_{0}}[{\bf p}%
-q{\bf A}({\bf r},t)]^{2}\psi ({\bf r},t)+q\Phi ({\bf r},t)\psi ({\bf r},t),%
\eqno
(1)
$$
where the operator ${\bf p}=-i\hbar \nabla $. The incorporation of the
vector potential corresponds to the understanding in classical mechanics
that the mechanical momentum in the Hamiltonian is given by the canonical
momentum ${\bf p}$ minus $q{\bf A}$.

It is known that in the presence of an external magnetic field, an
individual spectral line of atoms will split into a set of closely spaced
lines, known as the Zeeman splitting. The Zeeman effect has been accounted
for by using quantum mechanics. For this familiar case of interaction of a
hydrogen-like atom with the electromagnetic radiation from an external
source, Schr\"{o}dinger's equation becomes [4, 5] 
$$
i\hbar \frac{\partial }{\partial t}\psi =\frac{1}{2m_{0}}{\bf p}^{2}\psi
+q\Phi _{a}\psi -\frac{q}{m_{0}}{\bf A}\cdot {\bf p}\psi +\frac{1}{2m_{0}}(q%
{\bf A})^{2}\psi ,\eqno
(2)
$$
where $\Phi _{a}$ is the electric scalar (Coulomb) potential due to the
nucleus of the atom itself and ${\bf A}$ is the vector potential due to the
external source. As in the literature, we have made use of ${\bf p}\cdot (%
{\bf A}\psi )={\bf A}\cdot {\bf p}\psi $ since ${\bf p}\cdot {\bf A}$ is
taken to be zero. Physically, the Coulomb gauge $\nabla \cdot {\bf A}=0$
implies that the source generating the vector potential is electrically
neutralized, as it is ordinarily. If the term ${\bf p}\cdot {\bf A}$ does
not vanish, one has ${\bf p}\cdot {\bf A}/m_{0}=(i\hbar /m_{0}c^{2})\partial
\Phi /\partial t$ from the Lorenz gauge, where potential $\Phi $ is due to
the charge associated with the non-neutralized current generating potential $%
{\bf A}$. The angular frequency of an ordinary time-harmonic potential is
much lower than $m_{0}c^{2}/\hbar $. Therefore, the term ${\bf p}\cdot {\bf A%
}/m_{0}$ is much weaker than the corresponding potential $\Phi $ and hence
is not expected to have an appreciable physical consequence even if it does
exist. In our understanding, the reference frame of the position vector in
wavefunction and potentials and that of the time derivative are not
specified in Schr\"{o}dinger's equation.

Moreover, from the generalized Ehrenfest's theorem, the electromagnetic
force exerted on an unbounded charged particle due to the scalar and vector
potentials can be given in terms of the expectation values as [4] 
$$
{\bf F}=m_{0}\frac{d^{2}\left\langle {\bf r}\right\rangle }{dt^{2}}%
=-q\left\langle \nabla \Phi \right\rangle -q\left\langle \frac{\partial {\bf %
A}}{\partial t}\right\rangle +\frac{1}{2m_{0}}q\left\{ \left\langle ({\bf p}%
-q{\bf A})\times {\bf B}\right\rangle \ -\ \left\langle {\bf B}\times ({\bf p%
}-q{\bf A})\right\rangle \right\} ,\eqno
(3)
$$
where the magnetic flux density ${\bf B}=\nabla \times {\bf A}$. As noted in
[4], in this formula the operator ${\bf p}$ in $({\bf p}-q{\bf A})$ does not
commute with field ${\bf B}$. Thus the preceding force formula does not
agree exactly with the Lorentz force law. The discrepancy between these two
force formulas is expected to be $(i\hbar q/2m_{0})\nabla \times {\bf B}$.
Although this discrepancy may be small in magnitude, it has not yet solved
to our knowledge.

$\ $

\noindent {\large {\bf 3. Local-Ether Wave Equation and its Consequences}}

We then go on to consider the local-ether wave equation. It is postulated
that under the influence of the electric scalar potential $\Phi $, the
matter wave $\Psi $ of a charged particle is governed by the nonhomogeneous
wave equation proposed to be [1] 
$$
\left\{ \nabla ^{2}-\frac{1}{c^{2}}\frac{\partial ^{2}}{\partial t^{2}}%
\right\} \Psi ({\bf r},t)=\frac{\omega _{0}^{2}}{c^{2}}\left\{ 1+\frac{2}{%
\hbar \omega _{0}}q\Phi ({\bf r},t)\right\} \Psi ({\bf r},t),\eqno
(4)
$$
where the natural frequency $\omega _{0}$ as well as the charge $q$ is an
inherent constant of the particle, and the position vector ${\bf r}$ and the
time derivative are referred to the associated local-ether frame, which is
an ECI frame for earthbound particles.

If the potential $\Phi $ is weak, the wavefunction $\Psi $ tends to be close
to a space-time harmonic $e^{i{\bf k}\cdot {\bf r}}e^{-i\omega t}$, where $%
\omega ^{2}=\omega _{0}^{2}+c^{2}k^{2}$. Thus it has been shown that the
velocity of the particle is given by ${\bf v}={\bf k}c^{2}/\omega $ and
hence the angular frequency can be given by $\omega =\omega _{0}/\sqrt{%
1-v^{2}/c^{2}}$, where the velocity ${\bf v}$ is referred specifically to
the local-ether frame. As the natural frequency $\omega _{0}$ is shown to be
related to the rest mass $m_{0}$, the speed-dependent mass $m$ related to
the frequency $\omega $ is then given by the familiar form of $m=m_{0}/\sqrt{%
1-v^{2}/c^{2}}$ [1]. It is noted that these relations of the speed-dependent
angular frequency and wavelength of matter wave and of the speed-dependent
mass of particle look like the postulates of de Broglie and the Lorentz
mass-variation law, except that the particle speed $v$ is referred
specifically to the local-ether frame. Thereafter, by introducing the
reduced wavefunction $\psi $ given by $\Psi ({\bf r},t)=\psi ({\bf r},t)e^{i%
{\bf k}\cdot {\bf r}}e^{-i\omega t}$ and by expanding the Laplacian and the
time derivative in the d'Alembertian operator, it has been shown that the
preceding wave equation in $\Psi $ reduces to the first-order time evolution
equation in $\psi $ [1]. That is, 
$$
i\hbar \frac{\omega }{\omega _{0}}\frac{\partial }{\partial t}\psi ({\bf r}%
,t)=-\frac{\hbar ^{2}}{2m_{0}}\nabla ^{2}\psi ({\bf r},t)+q\Phi ({\bf r}%
,t)\psi ({\bf r},t)-i\frac{\hbar ^{2}}{m_{0}}{\bf k}\cdot \nabla \psi ({\bf r%
},t).\eqno
(5)
$$
Again, the position vector ${\bf r}$ here is referred to the local-ether
frame.

Consider a hydrogen-like atom which is moving at a velocity ${\bf v}_{a}$
with respect to the local-ether frame. It is expected that the electric
scalar potential $\Phi _{a}$ due to the nucleus of the atom will move with
this atom. Accordingly, this potential is stationary in the atom frame with
respect to which the atom is stationary, while it is moving in the
local-ether frame. Under Galilean transformations, the potential comoving
with the atom can be written as $\Phi _{a}({\bf r})$ or as $\Phi _{a}({\bf r}%
-{\bf v}_{a}t)$, where the position vector ${\bf r}$ is referred to the atom
or to the local-ether frame, respectively. The average value of the velocity
of the electron bounded in the atom should be identical to the atom velocity 
${\bf v}_{a}$; otherwise, the electron tends to escape from the atom. Thus
the spatial and temporal variation of the wavefunction of the bounded
electron can be expected to be close to the factored-out harmonic $e^{i{\bf k%
}\cdot {\bf r}}e^{-i\omega t}$ and then the reduced wavefunction is governed
by the preceding evolution equation, where the propagation vector ${\bf k}=m%
{\bf v}_{a}/\hbar $, the speed-dependent mass $m=m_{0}/\sqrt{%
1-v_{a}^{2}/c^{2}}$, the potential is given by $\Phi _{a}({\bf r}-{\bf v}%
_{a}t)$, and the position vector ${\bf r}$ is referred to the local-ether
frame.

Remark the Galilean transformation 
$$
\left( \frac{\partial f}{\partial t}\right) _{a}=\frac{\partial f}{\partial t%
}+{\bf v}_{a}\cdot \nabla f,\eqno
(6)
$$
where $\partial /\partial t$ and $(\partial /\partial t)_{a}$ denote the
time derivatives with respect to the local-ether and the atom frames and are
taken under constant ${\bf r}$ and (${\bf r}-{\bf v}_{a}t$), respectively,
as ${\bf r}$ is referred to the local-ether frame. Thereby, for the electron
bounded in the moving atom, the time evolution equation observed in the atom
frame becomes [1] 
$$
i\hbar \frac{\omega }{\omega _{0}}\frac{\partial }{\partial t}\psi ({\bf r}%
,t)=-\frac{\hbar ^{2}}{2m_{0}}\nabla ^{2}\psi ({\bf r},t)+q\Phi _{a}({\bf r}%
)\psi ({\bf r},t),\eqno
(7)
$$
where the position vector ${\bf r}$ and hence the time derivative are
referred to the atom frame, instead of the local-ether one. It is noted that
the time derivative connects with an extra multiplying term of $\omega
/\omega _{0}$ which is just the mass-variation factor $1/\sqrt{%
1-v_{a}^{2}/c^{2}}$. However, except for this factor, the time evolution
equation as well as the potential is independent of the motion of atom if
the atom frame is adopted as the reference frame. Consequently, the
solutions for the eigenfunction $\psi $ and the eigenvalue $\hbar \tilde{%
\omega}(\omega /\omega _{0})$ of this equation in the atom frame will be
independent of the atom speed $v_{a}$. Accordingly, as compared to that of a
stationary atom, the energy $\hbar \tilde{\omega}$ of each quantum state
will decrease with the inverse of the mass-variation factor when the atom is
moving at speed $v_{a}$ with respect to the local-ether frame.

The frequency of light emitted from or absorbed by an atom is known to be
equal to the transition frequency which in turn is proportional to the
difference in energy between two involved quantum states. Thus the
transition frequency $f$ is determined by the energy $\hbar \tilde{\omega}$
and then decreases with increasing atom speed by the inverse of the
mass-variation factor. That is, 
$$
f=f_{0}\sqrt{1-v_{a}^{2}/c^{2}},\eqno
(8)
$$
where the atom speed $v_{a}$ is referred specifically to the local-ether
frame and $f_{0}$ is the rest transition frequency of the atom when it is
stationary in this frame. Consequently, the transition frequency and hence
the clock rate of earthbound atomic clocks depend on earth's rotation, but
are entirely independent of earth's orbital motion. Thereby, the atomic
clock flying westward tends to have a lower speed and tick at a faster rate
than the one flying eastward. Thus the preceding formula accounts for the
east-west directional anisotropy in atomic clock rate demonstrated in the
Hafele-Keating experiment with circumnavigation clocks. On the other hand,
for a geostationary atom or an atom onboard an earth's satellite moving in a
circular orbit, the speed $v_{a}$ and hence the transition frequency $f$
remain unchanged with the passage of time. Thus the preceding formula also
accounts for the high synchronism among the various GPS atomic clocks moving
in nearly circular orbits and for the spatial isotropy in transition
frequency in the Hughes-Drever experiment with geostationary atoms [1].

Moreover, in order to derive the whole electromagnetic force, it is proposed
that the wave equation is modified by connecting the potential $\Phi $ to a
dimensionless operator $U$. For the electric scalar potential $\Phi $ due to
source particles of a given velocity ${\bf v}_{s}$ with respect to the
local-ether frame, it is postulated that the local-ether wave equation
incorporates the operator $U$ [3]. That is, 
$$
\left\{ \nabla ^{2}-\frac{1}{c^{2}}\frac{\partial ^{2}}{\partial t^{2}}%
\right\} \Psi {\bf (r},t)=\frac{\omega _{0}^{2}}{c^{2}}\left\{ 1+\frac{2}{%
\hbar \omega _{0}}q\Phi ({\bf r},t)(1+U)\right\} \Psi ({\bf r},t),\eqno
(9)
$$
where operator $U$ is derived from the Laplacian operator and is given by 
$$
U=\frac{1}{2c^{2}}\left( -i\frac{c^{2}}{\omega _{0}}\nabla -{\bf v}%
_{s}\right) ^{2}.\eqno
(10)
$$
The operator $U$ tends to enhance the effect of the electric scalar
potential and hence is called the augmentation operator. Again, the
local-ether wave equation can lead to a first-order time evolution equation
in terms of the reduced wavefunction $\psi $. When the particle speed is low
and then the propagation vector ${\bf k}$ in the factored-out harmonic is
taken to zero, the evolution equation reads 
$$
i\hbar \frac{\partial }{\partial t}\psi ({\bf r},t)=\frac{1}{2m_{0}}{\bf p}%
^{2}\psi ({\bf r},t)+q\Phi ({\bf r},t)(1+U)\psi ({\bf r},t),\eqno
(11)
$$
where the position vector ${\bf r}$ is referred to the local-ether frame. By
evaluating the velocity and then the acceleration of the charged particle
under the influence of the electric scalar potential connected to the
augmentation operator in a quantum-mechanically approach, the
electromagnetic force exerted on the particle has been derived [3].

Consider the force law for the ordinary case where the source particles are
drifting in a matrix and the ions which constitute the matrix tend to
electrically neutralize the mobile particles, such as electrons in a metal
wire. Suppose the neutralizing matrix is of an arbitrary charge density $%
\rho _{m}$ and moves as a whole at a velocity ${\bf v}_{m}$ with respect to
the local-ether frame, while the mobile source particles are of charge
density $\rho _{v}$ and move at ${\bf v}_{s}$ with respect to this frame.
Under the ordinary low-speed condition where all the involved particles move
slowly with respect to the local-ether frame and the sources drift very
slowly with respect to the matrix frame, the electromagnetic force law for a
particle of charge $q$ and inertial mass $m_{0}$ can be given in terms of
the local-ether potentials $\Phi $ and ${\bf A}$ [3]. That is, 
$$
{\bf F}({\bf r},t)=q\left\{ -\nabla \Phi ({\bf r},t)-\left( \frac{\partial }{%
\partial t}{\bf A}({\bf r},t)\right) _{m}+{\bf v}_{em}\times \nabla \times 
{\bf A}({\bf r},t)\right\} ,\eqno
(12)
$$
where the time derivative $(\partial /\partial t)_{m}$ is referred to the
matrix frame, the velocity difference ${\bf v}_{em}={\bf v}_{e}-{\bf v}_{m}$%
, and ${\bf v}_{e}$ is the velocity of the charged particle with respect to
the local-ether frame. The electric scalar potential $\Phi $ and the
magnetic vector potential ${\bf A}$ in turn are given by 
$$
\Phi ({\bf r},t)=\frac{1}{\epsilon _{0}}\int \frac{\rho _{n}({\bf r}^{\prime
},t-R/c)}{4\pi R}dv^{\prime }\eqno
(13)
$$
and 
$$
{\bf A}({\bf r},t)=\frac{1}{\epsilon _{0}c^{2}}\int \frac{{\bf J}_{n}({\bf r}%
^{\prime },t-R/c)}{4\pi R}dv^{\prime },\eqno
(14)
$$
where potential $\Phi $ is due to the net charge density $\rho _{n}=\rho
_{v}+\rho _{m}$, potential ${\bf A}$ is due to the neutralized current
density ${\bf J}_{n}={\bf v}_{sm}\rho _{v}$, ${\bf v}_{sm}$ ($={\bf v}_{s}-%
{\bf v}_{m}$) is the Newtonian relative velocity of the source particle with
respect to the matrix, and $R=|{\bf r}-{\bf r}^{\prime }|$. It has been
pointed out that the formula (12) is identical to the Lorentz force law, if
the latter is observed in the matrix frame, as done tacitly in common
practice [3].

$\ $

\noindent {\large {\bf 4. Modifications of Schr\"{o}dinger's Equation}}

In what follows, we derive from the local-ether wave equation the evolution
equation for the interaction of atoms with electromagnetic radiation.
Consider an atom which is moving at a velocity ${\bf v}_{a}$ with respect to
the local-ether frame. Again, the spatial variation of the wavefunction $%
\Psi $ of the bounded electron is expected to be close to the space harmonic 
$e^{i{\bf k}\cdot {\bf r}}$, where the position vector ${\bf r}$ is referred
to the local-ether frame. Thus the Laplacian becomes 
$$
\nabla ^{2}\Psi ({\bf r},t)=\left\{ \nabla ^{2}\tilde{\psi}({\bf r},t)+i2%
{\bf k}\cdot \nabla \tilde{\psi}({\bf r},t)-k^{2}\tilde{\psi}({\bf r}%
,t)\right\} e^{i{\bf k}\cdot {\bf r}},\eqno
(15)
$$
where $\Psi ({\bf r},t)=\tilde{\psi}({\bf r},t)e^{i{\bf k}\cdot {\bf r}}$
and $\tilde{\psi}$ is a weak function of space. The term associated with $%
k^{2}$ is neglected hereafter, as the atom speed $v_{a}$ is supposed to be
much lower than $c$. Thereby, the local-ether wave equation (9) becomes 
$$
\left\{ \nabla ^{2}-\frac{1}{c^{2}}\frac{\partial ^{2}}{\partial t^{2}}%
\right\} \tilde{\psi}({\bf r},t)=\frac{\omega _{0}^{2}}{c^{2}}\tilde{\psi}(%
{\bf r},t)+\frac{2\omega _{0}}{\hbar c^{2}}q\Phi (1+U_{k})\tilde{\psi}({\bf r%
},t)-i2{\bf k}\cdot \nabla \tilde{\psi}({\bf r},t),\eqno
(16)
$$
where the augmentation operator $U_{k}$ is defined as 
$$
U_{k}=\frac{1}{2c^{2}}\left( \frac{{\bf p}}{m_{0}}+\frac{\hbar {\bf k}}{m_{0}%
}-{\bf v}_{s}\right) ^{2}.\eqno
(17)
$$
It is seen that the incorporation of ${\bf k}$ is owing to the manipulation
that the space harmonic $e^{i{\bf k}\cdot {\bf r}}$ is factored out from $%
\Psi $.

Consider the ordinary case where the scalar potential $\Phi $ as well as the
spatial rate of variation of $\Psi $ is weak. Thus the temporal variation of 
$\Psi $ or $\tilde{\psi}$ is close to that of the harmonic $e^{-i\omega
_{0}t}$ and then the wavefunction can be given as $\tilde{\psi}({\bf r}%
,t)=\psi ({\bf r},t)e^{-i\omega _{0}t}$, where $\psi $ is a weak function of
space and time. Then its second time derivative becomes 
$$
\frac{\partial ^{2}}{\partial t^{2}}\tilde{\psi}({\bf r},t)=\left\{ \frac{%
\partial ^{2}}{\partial t^{2}}\psi ({\bf r},t)-i2\omega _{0}\frac{\partial }{%
\partial t}\psi ({\bf r},t)-\omega _{0}^{2}\psi ({\bf r},t)\right\}
e^{-i\omega _{0}t}.\eqno
(18)
$$
As the temporal variation of $\psi $ is relatively weak, the second
derivative $\partial ^{2}\psi /\partial t^{2}$ can be neglected. Then we
have the first-order time evolution equation in terms of the reduced
wavefunction $\psi $ 
$$
\frac{\partial }{\partial t}\psi ({\bf r},t)=i\frac{c^{2}}{2\omega _{0}}%
\nabla ^{2}\psi ({\bf r},t)-i\frac{1}{\hbar }q\Phi (1+U_{k})\psi ({\bf r},t)-%
\frac{c^{2}}{\omega _{0}}{\bf k}\cdot \nabla \psi ({\bf r},t).\eqno
(19)
$$
We then go on to rearrange the evolution equation to express it in the atom
frame, instead of the local-ether frame. The last term in the preceding
equation can be written as $-{\bf v}_{a}\cdot \nabla \psi $, since the
propagation vector can be given by ${\bf k}=m_{0}{\bf v}_{a}/\hbar $. Thus,
by using the Galilean transformation (6) again, the time evolution equation
for the electron bounded in a moving atom becomes 
$$
i\hbar \frac{\partial }{\partial t}\psi ({\bf r},t)=-\frac{\hbar ^{2}}{2m_{0}%
}\nabla ^{2}\psi ({\bf r},t)+q\Phi ({\bf r},t)(1+U_{k})\psi ({\bf r},t),%
\eqno
(20)
$$
where the position vector ${\bf r}$ and hence the time derivative are
referred to the atom frame. As the potential in the preceding equation is
given simply by $\Phi ({\bf r},t)$ with ${\bf r}$ being referred to the atom
frame, it has to be given by $\Phi ({\bf r}-{\bf v}_{a}t,t)$ in (19) with $%
{\bf r}$ being referred to the local-ether frame.

Ordinarily, the electromagnetic radiation in the interaction of atoms comes
from a neutralized source. The electric scalar potential is supposed to be
composed as $\Phi =\Phi _{a}+\Phi _{s}+\Phi _{m}$, where $\Phi _{a}$ is due
to the nucleus of the atom, $\Phi _{s}$ to the mobile charged particles
forming the current in an external source, and $\Phi _{m}$ to the matrix in
this neutralized source. Under complete neutralization, $\Phi _{s}=-\Phi _{m}
$. Further, suppose the source is stationary in the atom frame, that is, $%
{\bf v}_{m}={\bf v}_{a}$. Thus the Doppler frequency shift is not involved.
Then it is easy to show that 
$$
(\Phi _{s}+\Phi _{m})(1+U_{k})=\frac{\Phi _{s}}{2c^{2}}\left\{ \left( \frac{%
{\bf p}}{m_{0}}-{\bf v}_{sm}\right) ^{2}-\left( \frac{{\bf p}}{m_{0}}\right)
^{2}\right\} =-\frac{1}{m_{0}}{\bf A}\cdot {\bf p}+\frac{1}{2}{\bf A}\cdot 
{\bf v}_{sm},\eqno
(21)
$$
where the vector potential due to the neutralized source is given according
to (14) as 
$$
{\bf A}=\frac{1}{c^{2}}{\bf v}_{sm}\Phi _{s}.\eqno
(22)
$$
Thus the atom-frame evolution equation becomes 
$$
i\hbar \frac{\partial }{\partial t}\psi ({\bf r},t)=\frac{1}{2m_{0}}{\bf p}%
^{2}\psi ({\bf r},t)+q\Phi _{a}({\bf r})\psi ({\bf r},t)-\frac{q}{m_{0}}{\bf %
A}\cdot {\bf p}\psi +\frac{1}{2}q{\bf A}\cdot {\bf v}_{sm}\psi ,\eqno
(23)
$$
where the augmentation operator connected to potential $\Phi _{a}$ is
neglected as the effect of its influence on the quantum states is small in
the interaction. Again, the potential $\Phi _{a}$ is supposed to move with
the atom and is stationary in the atom frame, while it is moving in the
local-ether frame. It is noted that the time evolution equation as well as
the potential is independent of the motion of atom if the atom frame is
adopted as the reference frame.

The preceding evolution equation then looks like Schr\"{o}dinger's equation
(2), except the last term and the reference frame. The last term in the
preceding equation is of the second order of normalized speed $v_{sm}/c$ and
is very weak as the drift speed $v_{sm}$ is very low, while the
corresponding one in Schr\"{o}dinger's equation is a quadratic term of $q%
{\bf A}$. It is seen that the latter is smaller in magnitude than the former
by a factor of $|q\Phi _{s}|/m_{0}c^{2}$ which in turn is much less than
unity. Anyway, this second-order interaction is commonly ignored in analysis
[4, 5] and no quantitative measurements are reported, to our knowledge.
However, one fundamental difference is that the position vector, the time
derivative, and the drift velocity are referred specifically to the atom
frame. In the perturbational treatment of interaction, the quantum states of
the unperturbed system are usually taken from the solutions of
Schr\"{o}dinger's equation with a {\it stationary} potential $\Phi _{a}$. By
so doing, one has actually adopted the atom frame as the reference frame
tacitly, although the result can be frame-independent. Thereby, {\bf the
preceding evolution equation is identical to Schr\"{o}dinger's equation, if
the latter is observed in the atom frame} as done tacitly in common
practice. In other words, Schr\"{o}dinger's equation (2) with a stationary
potential $\Phi _{a}$ has some hidden restrictions. That is, the reference
frame is actually the atom frame, the atom speed is low in the local-ether
frame, the source generating the interacting potential is electrically
neutralized and is stationary in the atom frame, and the drift speed in the
source is very low in this frame. However, these conditions are so common as
to be ignored easily.

The evolution equation (11), (20), or (23) then presents modifications of
Schr\"{o}dinger equation, referred specifically either to the local-ether or
to the atom frame. It is noted that the atom-frame evolution equation (23)
is independent of the atom velocity itself. Therefore, its consequences
comply with Galilean relativity and hence are independent of earth's
motions. However, this is owing to the approximation that the $k^{2}$ term
in (15) is omitted. When this second-order term is retained and hence the
restriction of low atom speed is removed, the effect of earth's rotation
resumes as in the consequences of (7).

$\ $

\noindent {\large {\bf 5. Conclusion}}

Based on the local-ether wave equation incorporating a nature frequency, the
electric scalar potential, and the augmentation operator, the first-order
time evolution equation is derived for a harmonic-like wavefunction. Then
the effect of the augmentation operator in the interaction of atoms with
electromagnetic radiation is discussed. Except a small second-order term,
this evolution equation can be identical to Schr\"{o}dinger's equation with
the vector potential, if the latter is observed in the atom frame. In common
practice, this frame has been adopted tacitly as the reference frame, as the
scalar potential due to the atom is taken to be stationary. The predicted
second-order interaction is stronger than the one in Schr\"{o}dinger's
equation and might provide a means to test the modified equation.

Besides, the local-ether wave equation leads to the speed-dependences in the
angular frequency and wavelength of matter wave, in the mass of particle,
and in the energy of quantum states. Thus it provides the physical origin of
the postulates of de Broglie and the Lorentz mass-variation law. Moreover,
it accounts for the east-west directional anisotropy in the Hafele-Keating
experiment, the synchronism in GPS, and for the spatial isotropy in the
Hughes-Drever experiment. It also leads to the electromagnetic force law in
conjunction with the physical origin of the inertial mass. Thus the
local-ether wave equation and the modified Schr\"{o}dinger equation account
for a variety of phenomena in a consistent way.

$\ $

$\ $

\noindent {\large {\bf References}}

\begin{itemize}
\item[{\lbrack 1]}]  C.C. Su, {\it Eur. Phys. J. B} {\bf 24}, 231 (2001).

\item[{\lbrack 2]}]  C.C. Su, {\it Eur. Phys. J. C} {\bf 21}, 701 (2001); 
{\it Europhys. Lett}. {\bf 56}, 170 (2001).

\item[{\lbrack 3]}]  C.C. Su, ``A local-ether wave\ equation and the
consequent electromagnetic force law,'' {\it J. Electromagnetic Waves
Applicat.} (in press); in {{\it IEEE Antennas Propagat. Soc. Int}. {\it Symp}%
. {\it Dig.}} (2001), vol. 1, p. 216; in{\ {\it Bull. Am. Phys. Soc.}} (Mar.
2{001}){\it ,} p. 1144{.}

\item[{\lbrack 4]}]  L.I. Schiff, {\it Quantum Mechanics} (McGraw-Hill, New
York, 1968), sects. 24 and 44.

\item[{\lbrack 5]}]  E. Merzbacher, {\it Quantum Mechanics} (Wiley, New
York, 1998), ch. 19.
\end{itemize}

\end{document}